\documentclass[sn-nature]{sn-jnl}

\usepackage{graphicx}%
\usepackage{multirow}%
\usepackage{amsmath,amssymb,amsfonts}%
\usepackage{amsthm}%
\usepackage{mathrsfs}%
\usepackage[title]{appendix}%
\usepackage{xcolor}%
\usepackage{textcomp}%
\usepackage{manyfoot}%
\usepackage{booktabs}%
\usepackage{algorithm}%
\usepackage{algorithmicx}%
\usepackage{algpseudocode}%
\usepackage{listings}%
\usepackage{caption}

\usepackage{amsmath}
\setcitestyle{super}

\definecolor{fuchsia}{rgb}{0.54, 0.17, 0.89}
\definecolor{azure}{rgb}{0.0, 0.5, 1.0}
\definecolor{pgreen}{rgb}{0.12, 0.3, 0.17}
\definecolor{alizarin}{rgb}{0.82, 0.1, 0.26}

\newcommand{\oiii}{[\textrm{O}~\textsc{iii}]}
\newcommand{\oiiir}{$[\textrm{O}~\textsc{iii}]_{\lambda5008}$}

\newcommand{\nii}{[\textrm{N}~\textsc{ii}]}

\newcommand{\ciii}{\textrm{C}~\textsc{iii}]}
\newcommand{\civ}{\textrm{C}~\textsc{iv}}

\newcommand{\hei}{\textrm{He}~\textsc{i}}
\newcommand{\heii}{\textrm{He}~\textsc{ii}}

\newcommand{\heiiuv}{$\textrm{He}~\textsc{ii}_{\lambda1640}$}
\newcommand{\heiiopt}{$\textrm{He}~\textsc{ii}_{\lambda4686}$}

\newcommand{\simgt}{\,\rlap{\lower 3.5 pt \hbox{$\mathchar \sim$}} \raise
1pt \hbox {$>$}\,}
\newcommand{\simlt}{\,\rlap{\lower 3.5 pt \hbox{$\mathchar \sim$}} \raise
1pt \hbox {$<$}\,}

\newcommand{\logoh}{$\log {\rm (O/H)}$}

\newcommand{\ha}{${\rm H\alpha}$}
\newcommand{\hb}{${\rm H\beta}$}

\newcommand{\jwst}{{JWST}}

\newcommand{\cl}{{\tt Cloudy}}

\newcommand{\magnifberr}{$71.3_{-8.3}^{+18.4}$}
\newcommand{\rsrcerr}{$36_{-8}^{+9}$}

\newcommand{\heiiuvsigma}{{$23.3\,\sigma$}}
\newcommand{\heiioptsigma}{{$7.7\,\sigma$}}

\newcommand{\id}{AMORE6} 

\newcommand{\idb}{\id-B}
\newcommand{\zspec}{5.725}
\newcommand{\zspecprev}{5.725}

\newcommand{\oxygensunerr}{$0.14\pm0.02\,\%$}

\newcommand{\oxygenerr}{{5.8_{-0.1}^{+0.1}}}

\newcommand{\logohdirecterr}{$6.03^{+0.10}_{-0.09}$}
\newcommand{\oxygensundirect}{{$0.22\,\%$}}

\newcommand{\oxygensundirectabsterr}{{$2.2\pm0.5\times10^{-3}$}}

\newcommand{\snlim}{3}
\newcommand{\scalehb}{1.2}

\newcommand{\dateobs}{{November {16--24}, 2026}}
\newcommand{\vermsaexp}{{ver.0.9.20}}

\newcommand{\pmap}{{1584}}

\theoremstyle{thmstyleone}%
\theoremstyle{thmstyletwo}%

\theoremstyle{thmstylethree}%

\begin{document}

\title[]{Nearly pristine gas and hard ionizing radiation one billion years after the Big Bang }


\author*[1]{\fnm{Takahiro} \sur{Morishita}}\email{morishita@astr.tohoku.ac.jp}
\author[2,3]{\fnm{Massimo} \sur{Stiavelli}}
\author[4]{\fnm{Tommaso} \sur{Treu}}
\author[5]{\fnm{Adriano} \sur{Fontana}}
\author[6]{\fnm{Guido} \sur{Roberts-Borsani}}
\author[6]{\fnm{Raunaq Singh} \sur{Rai}}
\author[7]{\fnm{Giorgia} \sur{Di Rosa}}
\author[8]{\fnm{Marco} \sur{Lombardi}}
\author[7]{\fnm{Piero} \sur{Rosati}}
\author[5]{\fnm{Borja} \sur{P\'{e}rez-D\'{i}az}}
\author[9]{\fnm{Eros} \sur{Vanzella}}
\author[10]{\fnm{Karl} \sur{Glazebrook}}
\author[11]{\fnm{Themiya} \sur{Nanayakkara}}
\author[12]{\fnm{Ana} \sur{Acebron}}
\author[13]{\fnm{Lorenzo} \sur{Bazzanini}}
\author[9]{\fnm{Pietro} \sur{Bergamini}}
\author[5]{\fnm{Davide} \sur{Bevacqua}}
\author[14,15]{\fnm{Maru\v{s}a} \sur{Brada\v{c}}}
\author[5]{\fnm{Antonello} \sur{Calabr\`{o}}}
\author[16]{\fnm{Riccardo} \sur{Caleno}}
\author[5]{\fnm{Marco} \sur{Castellano}}
\author[16,1]{\fnm{Sunmyon} \sur{Chon}}
\author[6]{\fnm{Richard S} \sur{Ellis}}
\author[17]{\fnm{Elena} \sur{Gallo}}
\author[5]{\fnm{Giovanni} \sur{Gandolfi}}
\author[16]{\fnm{Luca} \sur{Graziani}}
\author[8,18]{\fnm{Claudio} \sur{Grillo}}
\author[6]{\fnm{Joe} \sur{Hopkins}}
\author[15]{\fnm{Tucker} \sur{Jones}}
\author[19]{\fnm{Zhaoran} \sur{Liu}}
\author[5]{\fnm{Mario} \sur{Llerena}}
\author[4]{\fnm{Matthew} \sur{Malkan}}
\author[9]{\fnm{Massimo} \sur{Meneghetti}}
\author[9]{\fnm{Matteo} \sur{Messa}}
\author[5]{\fnm{Lorenzo} \sur{Napolitano}}
\author[20]{\fnm{Evan Haze} \sur{Nunez-Cravin}}
\author[21]{\fnm{Lorenzo} \sur{Parente}}
\author[4]{\fnm{Massimo} \sur{Pascale}}
\author[5]{\fnm{Laura} \sur{Pentericci}}
\author[22]{\fnm{Giulia} \sur{Rodighiero}}
\author[16,23]{\fnm{Matteo} \sur{Saggini}}
\author[5]{\fnm{Paola} \sur{Santini}}
\author[16]{\fnm{Raffaella} \sur{Schneider}}
\author[24]{\fnm{Michele} \sur{Trenti}}
\author[25]{\fnm{Benedetta} \sur{Vulcani}}
\author[16,5]{\fnm{Tommaso} \sur{Zana}}
\author[17]{\fnm{Fan} \sur{Zou}}

\affil*[1]{\orgdiv{Astronomical Institute, Graduate School of Science}, \orgname{Tohoku University}, \orgaddress{\city{Sendai}, \country{Japan}}}
\affil[2]{\orgname{Space Telescope Science Institute}, \orgaddress{\city{Baltimore}, \country{USA}}}
\affil[3]{\orgname{The William H. Miller III Dept. of Physics and Astronomy}, \orgaddress{\city{Baltimore}, \country{USA}}}
\affil[4]{\orgdiv{Department of Physics and Astronomy}, \orgname{University of California, Los Angeles}, \orgaddress{\city{Los Angeles}, \country{USA}}}
\affil[5]{\orgdiv{Osservatorio Astronomico di Roma}, \orgname{INAF}, \orgaddress{\city{Monte Porzio Catone}, \country{Italy}}}
\affil[6]{\orgdiv{Department of Physics and Astronomy}, \orgname{University College London}, \orgaddress{\city{London}, \country{UK}}}
\affil[7]{\orgdiv{Dipartimento di Fisica e Scienze della Terra}, \orgname{Universit\`{a} degli Studi di Ferrara}, \orgaddress{\city{Ferrara}, \country{Italy}}}
\affil[8]{\orgdiv{Dipartimento di Fisica ``Aldo Pontremoli''}, \orgname{Universit\`{a} degli Studi di Milano}, \orgaddress{\city{Milano}, \country{Italy}}}
\affil[9]{\orgdiv{Osservatorio di Astrofisica e Scienza dello Spazio di Bologna}, \orgname{INAF}, \orgaddress{\city{Bologna}, \country{Italy}}}
\affil[10]{\orgdiv{Centre for Astrophysics and Supercomputing}, \orgname{Swinburne University of Technology}, \orgaddress{\city{Hawthorn}, \country{Australia}}}
\affil[11]{\orgdiv{Sydney Institute for Astronomy, School of Physics}, \orgname{The University of Sydney}, \orgaddress{\city{Sydney}, \country{Australia}}}
\affil[12]{\orgname{Instituto de F\'isica de Cantabria}, \orgaddress{\city{Santander}, \country{Spain}}}
\affil[13]{\orgdiv{Istituto di Radioastronomia}, \orgname{INAF}, \orgaddress{\city{Bologna}, \country{Italy}}}
\affil[14]{\orgdiv{Faculty of Mathematics and Physics}, \orgname{University of Ljubljana}, \orgaddress{\city{Ljubljana}, \country{Slovenia}}}
\affil[15]{\orgdiv{Department of Physics and Astronomy}, \orgname{University of California, Davis}, \orgaddress{\city{Davis}, \country{USA}}}
\affil[16]{\orgdiv{Dipartimento di Fisica}, \orgname{Sapienza Universit\`{a} di Roma}, \orgaddress{\city{Roma}, \country{Italy}}}
\affil[17]{\orgdiv{Department of Astronomy}, \orgname{University of Michigan}, \orgaddress{\city{Ann Arbor}, \country{USA}}}
\affil[18]{\orgdiv{IASF Milano}, \orgname{INAF}, \orgaddress{\city{Milano}, \country{Italy}}}
\affil[19]{\orgdiv{Institute for Astronomy}, \orgname{University of Hawaii}, \orgaddress{\city{Honolulu}, \country{USA}}}
\affil[20]{\orgdiv{The Observatories}, \orgname{Carnegie Institution for Science}, \orgaddress{\city{Pasadena}, \country{USA}}}
\affil[21]{\orgdiv{Dipartimento di Fisica e Astronomia}, \orgname{Universit\`{a} di Bologna}, \orgaddress{\city{Bologna}, \country{Italy}}}
\affil[22]{\orgname{Dipartimento di Fisica e Astronomia}, \orgaddress{\city{Padova}, \country{Italy}}}
\affil[23]{\orgdiv{Dipartimento di Fisica}, \orgname{Universit\`{a} di Roma Tor Vergata}, \orgaddress{\city{Roma}, \country{Italy}}}
\affil[24]{\orgdiv{School of Physics}, \orgname{The University of Melbourne}, \orgaddress{\city{Melbourne}, \country{Australia}}}
\affil[25]{\orgdiv{Osservatorio Astronomico di Padova}, \orgname{INAF}, \orgaddress{\city{Padova}, \country{Italy}}}

\abstract{
The first stars formed out of light elements (H, He, and Li leftover from Big Bang Nucleosynthesis) and began the chemical enrichment of the Universe. Identifying this first-generation population of stars, known as Population III, and understanding their properties is one of the most ambitious goals of modern extragalactic astrophysics. We report the detection of He and H emission lines and extremely low Oxygen abundance in \id, a strongly magnified galaxy at redshift $z=\zspec$, approximately one billion years after the Big Bang, with deep James Webb Space Telescope spectroscopy. He\,\textsc{ii} $\lambda1640$ and $\lambda4686$ emission lines are detected at \heiiuvsigma\ and \heiioptsigma\ significance, respectively. The observed He\,\textsc{ii}-to-H$\beta$ ratios are consistent with predictions for the extreme ionizing radiation fields and the low chemical compositions typical of Population III stars. Oxygen abundance of the interstellar medium is estimated to be \oxygensundirectabsterr\ solar from weak [OIII]$\lambda4363$ and $\lambda5008$ lines, consistent with an almost pristine environment, perhaps enriched by the Pop III stars themselves.  
Although we cannot exclude alternative sources of hard ionizing radiation, such as accreting intermediate-mass black holes, these observations offer the first potential view of the extremely low metallicity environment where the Universe's first stars are predicted to be born. The data show that such pockets of pristine gas still exist one billion years after the Big Bang and that they are dominated by very massive stars.
}
\maketitle

The observational properties of the very first generation of stars formed in the Universe (the so-called Pop III stars, \cite{bromm03,wise2012, Klessen2023, Venditti2026, Katz2026}) remain poorly constrained given the challenge of isolating their contributions from the integrated light of distant galaxies \citep{maiolino26}. Spectroscopy with the James Webb Space Telescope (JWST) has revealed substantial chemical enrichment in galaxies as far out as $z\gtrsim10$, extending to epochs only a few hundred million years after the Big Bang \citep{curtis-lake23,castellano24,carniani24,naidu25,rb26}. The search for primordial star formation need not be confined to the highest redshifts: small, relatively isolated systems may form their first generation of stars from pristine gas even several hundred million years after the Big Bang \citep{stiavelli2010,trenti2009, Tornatore2007, Venditti2023}. From an observational point of view, a distinct feature of Pop III stars is a strong nebular He\,\textsc{ii} emission due to the hard ionizing radiation produced by extremely metal-poor or metal-free stars \citep{schaerer02,nakajima22,Venditti2026,maiolino26}, in combination with the absence of emission lines from heavier elements, which have not formed yet in appreciable amount in such systems.

The strongly lensed source AMORE6, residing behind the Abell 2744 galaxy cluster, has been identified as a primary target to search for Pop III stars. At $z_{\rm spec}=\zspecprev$, its two lensed images are magnified by factors of several tens, making this intrinsically faint source accessible to spectroscopic characterization. Earlier JWST/NIRCam slitless spectroscopy detected \hb\ but not \oiii$_{\lambda\lambda4960,5008}$ in either image. The application of Oxygen estimators based on strong nebular lines to those data implied an upper limit of $12+\log({\rm O/H})<5.97$ (3$\sigma$), or approximately $<0.19\%$ of the solar oxygen abundance \citep{morishita25}. The source is extremely faint (absolute UV magnitude $M_{\rm UV}=-14.5\pm0.1$), blue (UV spectral slope $\beta_{\rm UV}=-2.8\pm0.1$), and compact (effective radius $R_e=$\,\rsrcerr\,pc). Together, these properties made AMORE6 a promising candidate site for a young burst of star formation in minimally enriched gas, but definitive proof was lacking.

\begin{figure*}
\centering
	\includegraphics[width=0.95\textwidth]{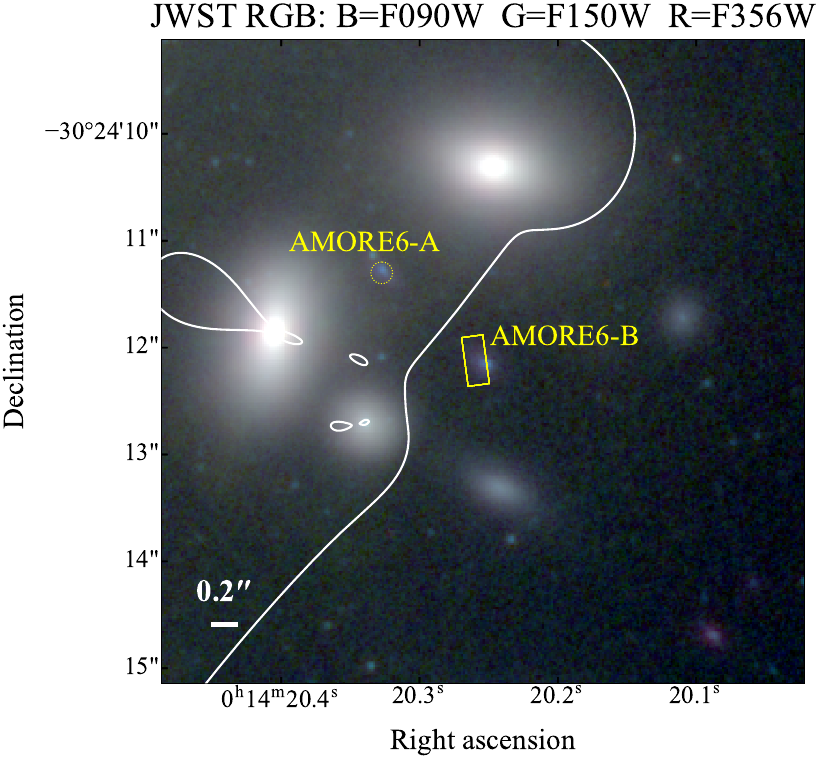}
	\caption{
    False-color NIRCam images of \id-B, constructed from images taken in the F090W (blue), F150W (green) and F356W (red) filters. North is up. The NIRSpec MSA slit position is overlaid (yellow rectangle). The white contour indicates the critical curve, where gravitational lensing produces high magnification, predicted from a gravitational lens model of this cluster \cite{bergamini23b}. The position of \id-A is also indicated by a yellow circle.
    }
\label{fig:mosaic}
\end{figure*}

\begin{figure}
\centering
	\includegraphics[width=0.99\textwidth]{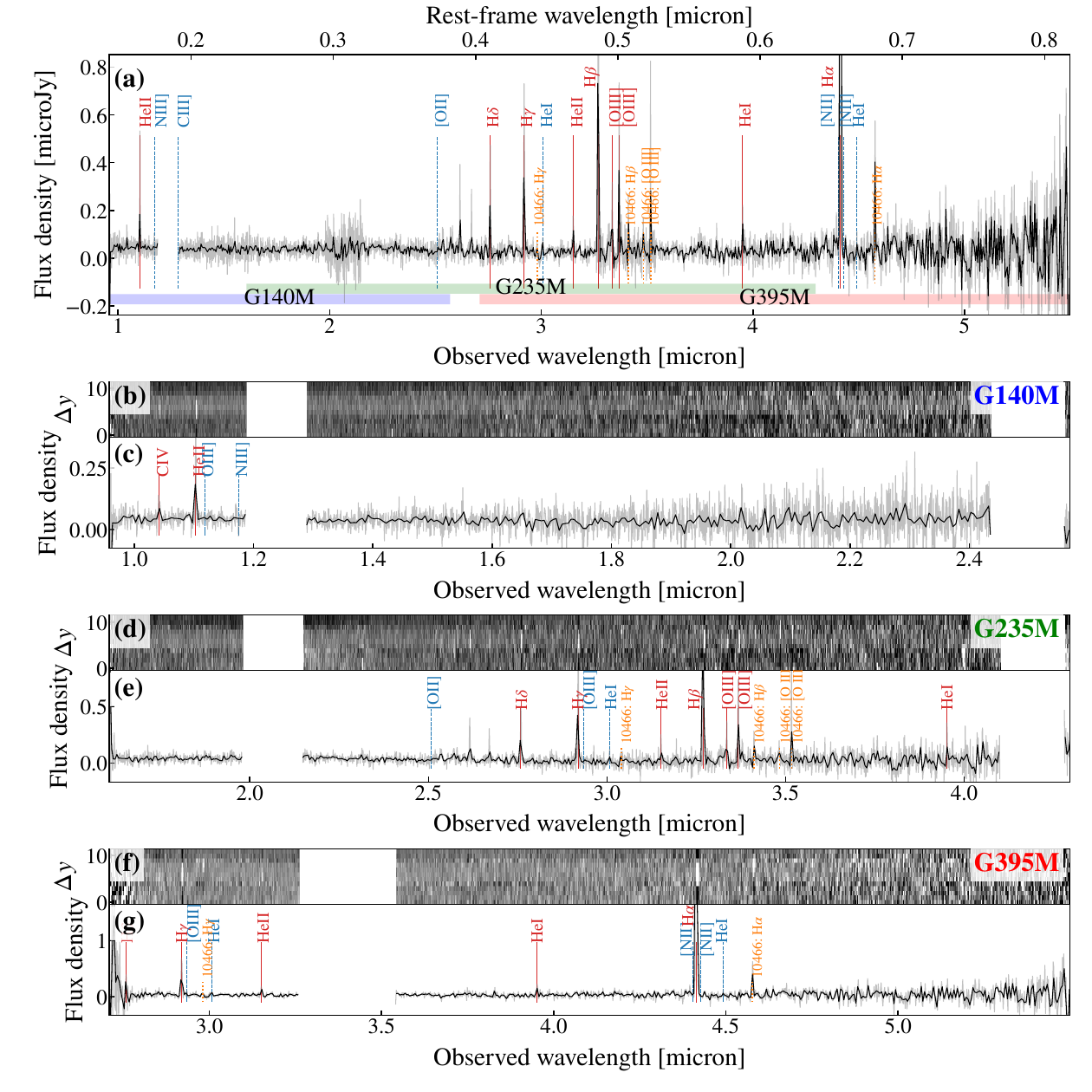}
	\caption{
    JWST/NIRSpec spectroscopy of \id. \textbf{(a)} Combined one-dimensional (1D) spectrum obtained with the G140M, G235M, and G395M gratings, covering an observed wavelength range of $\sim1$--$5.5\,\mu$m. Black solid line shows the binned spectrum by $50$\,\AA\ and gray thin line the original spectrum. 
    The upper axis indicates the rest-frame wavelength. Vertical lines mark the expected positions of emission lines at the redshift of \id. Red solid lines indicate those detected and blue dashed lines non-detected. 
    Orange dotted lines mark emission lines of the contaminant 10466 at $z=6.18$ (Methods). 
    The colored horizontal bars indicate the wavelength coverage of the three gratings. \textbf{(b), (c)} Two-dimensional (2D) and one-dimensional (1D) spectra obtained with G140M. \textbf{(d), (e)} Same as panels (b) and (c) but for G235M. \textbf{(f), (g)} Same as panels (b) and (c) but for G395M. 
    }
\label{fig:spec_all}
\end{figure}

\begin{figure}
\centering
	\includegraphics[width=0.99\textwidth]{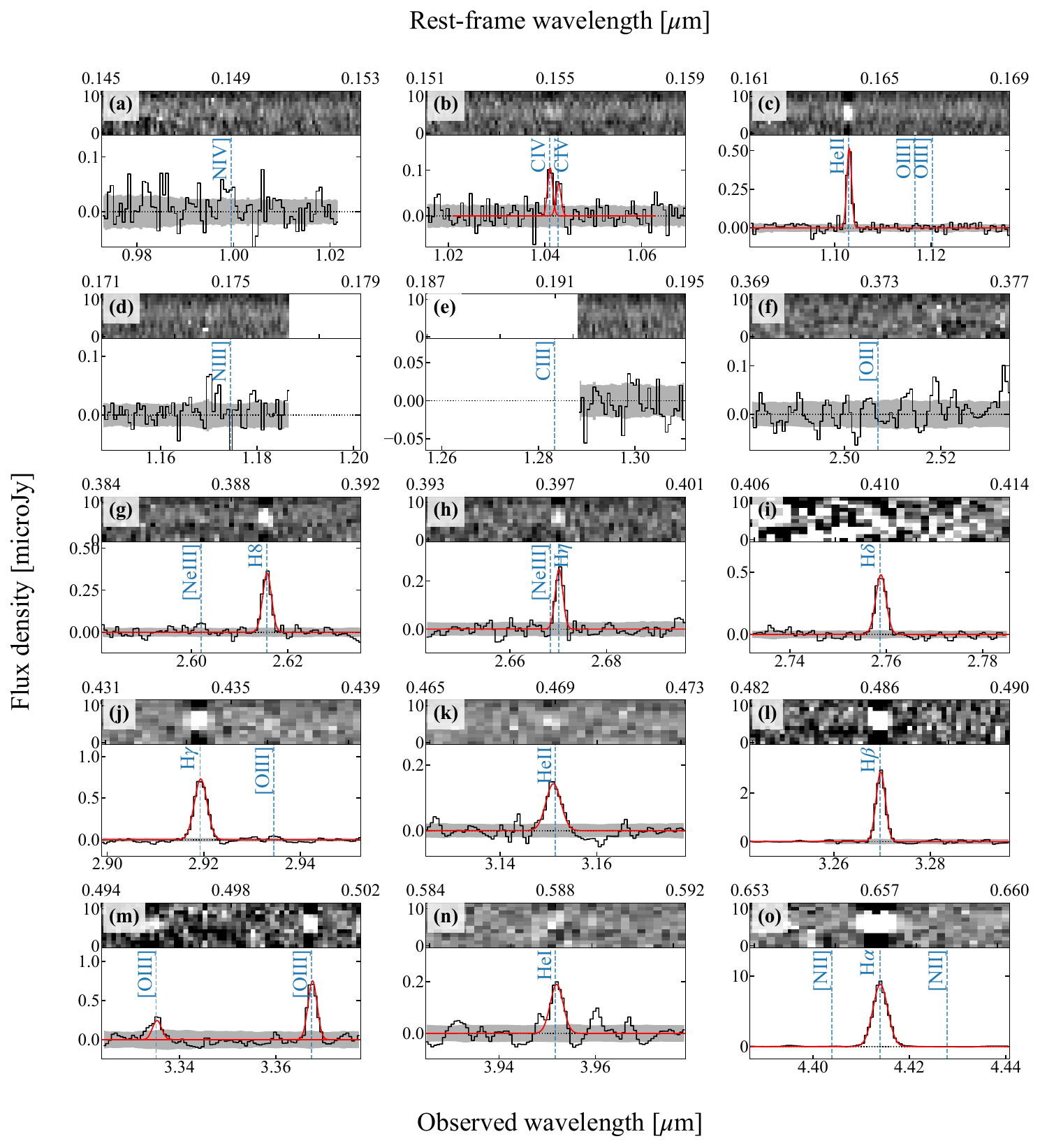}
	\caption{
    Snapshots of the JWST/NIRSpec spectrum of AMORE6 centred at the position of selected rest-frame lines  obtained with
    JWST/NIRSpec.
    The black solid lines show the observed spectra, while the
    gray shaded regions represent the $1\,\sigma$ uncertainties.
    Vertical blue dashed lines indicate the expected wavelengths
    of individual emission lines at the systemic redshift of \id.
    Red curves indicate Gaussian fits where a line is considered detected in the current analysis. We note that local continuum is subtracted in each wavelength window.
    }
\label{fig:spec_lines}
\end{figure}

We report deep JWST/NIRSpec multi-object spectroscopic follow up of the brighter lensed image, AMORE6-B (Fig.~\ref{fig:mosaic}), obtained as part of the Deep Amplified JWST Experiment (DAJE) program (GO 9645; PIs: T.~Treu, A.~Fontana, G.~Roberts-Borsani). The three medium-resolution ($R\sim1000$) gratings, G140M ($\sim50.8$ hours on-source), G235M ($\sim45.2$ hours) and G395M ($\sim38.2$ hours), provide nearly continuous coverage from approximately 1 to 5.5 $\mu$m (Fig.~\ref{fig:spec_all}), encompassing rest-frame ultraviolet and optical emission. We detect He\,\textsc{ii} $\lambda1640$ and $\lambda4686$ emission at the significance of \heiiuvsigma\ and \heiioptsigma, respectively, and hydrogen recombination lines including H$\delta$, H$\gamma$, H$\beta$ and H$\alpha$, and weak \oiiir\ {and \civ}. We do not conclusively detect emission from any other elements (signal-to-noise ratio, SNR\,$<3$), as shown in  Fig.~\ref{fig:spec_lines}. Individual flux measurements, detection thresholds, extinction correction, and wavelength-dependent slit losses are described in the Supplementary Information section.

The detection of both \heii\ recombination lines with large intensities and equivalent widths provides direct evidence for an exceptionally hard ionizing radiation field. We measure rest-frame equivalent widths of $22\pm2$\,\AA\ and $30\pm6$\,\AA\ for \heiiuv\ and \heii$_{\lambda4686}$, respectively. The production of nebular \heii\ emission requires photons with energies above 54.4\,eV, which are difficult to produce in sufficient quantities to explain the observed line strengths with conventional stellar populations. Such a radiation field is, instead, a natural outcome of star formation out of nearly pristine gas, where the initial mass function (IMF) is expected to be top-heavy due to the lack of efficient metal fine-structure line cooling, leading to the formation of very massive stars \citep{Bromm2003, Omukai2005, Schneider2002, Chon2021}. These stars burn hydrogen through the CNO cycle, reaching temperatures as high as $10^5$~K (as opposite to the $4\times10^4$~K of ordinary massive stars), thus emitting photons above 54.4\,eV. The absence of metal-line blanketing and the weakness of radiatively driven stellar winds further enhance their emergent extreme-ultraviolet radiation. Both the measured equivalent widths and the \heii-to-\hb\ flux ratios are consistent with predictions for metal-free or extremely metal-poor stellar populations (Fig.~\ref{fig:he2}; see Supplementary Information), suggesting that very hot, massive stars may be responsible for the exceptionally hard ionizing radiation field in \id.

{Alternatively, or in addition, the hard ionizing radiation could plausibly originate from an accreting black hole.} If the emission is powered by accretion, empirical scaling relations\citep{2013ARA&A..51..511K,2020ARA&A..58..257G} suggest a black hole mass of $M_{\rm BH}\sim10^{3}$--$10^{4}\,M_\odot$, implying an intermediate-mass black hole (IMBH), a possibility of particular interest in such a chemically primitive system. A similar black hole mass $\sim10^4\ M_\odot$ is obtained from the required hard ionizing continuum, a power-law spectrum and Eddington luminosity.
The medium-resolution JWST/NIRSpec spectroscopy does not provide a definitive kinematic test for such a low-mass black hole. 
Thus, the absence of a broad component does not exclude the IMBH scenario. Given an IMBH cannot be excluded, the combination of strong \heii\ emission and stringent metal-line limits makes AMORE6 an exceptional laboratory for investigating primordial stellar populations and/or early black hole formation.

The \oiii\ emission supports an exceptionally low oxygen abundance. Using the observed \hb\ and \oiii\ line fluxes and photoionization models, we infer an oxygen abundance of $12+\log({\rm O/H})=$\,\logohdirecterr, corresponding to \oxygensundirectabsterr\ of the solar value (Methods). This stringent estimate remains largely unchanged under different methods or assumptions about the ionization conditions, including a lower ionization parameter and the contribution of higher ionization states of oxygen.

Collisional de-excitation at high electron densities can suppress optical forbidden lines and mimic low metallicity. 
We have explored these options utilising the additional constraints that can be derived from  the faint detections in \oiiir\ and \civ\ that we observe in the spectra as well as the He~II line detections.
Alternative interpretations to low metallicity, involving gas density, ionization parameter, elemental abundance ratios, and ionizing spectrum are discussed in the Supplementary Information and Extended Data Section. \textbf{Fig.~\ref{fig:line_ratios}.} In short, while the inferred abundance depends on the nebular models, the combined metal-line limits favor an environment with low chemical enrichment.

The size and mass of AMORE6 also inform its classification. Its compact morphology and low stellar mass are consistent with a recent and intense episode of star formation, although a young massive star cluster is also possible \citep{vanzella23}. Its effective radius measured in the source plane ($R_e=$\,\rsrcerr\,pc\cite{morishita25}, corrected for magnification) places it near the locus of massive star clusters and ultra-compact dwarfs \citep{morishita26a}. Its relatively smooth morphology and size, larger than those of typical large massive star clusters, favor a galactic interpretation (Extended Data Fig.~\ref{fig:slit}). Moreover, its inferred stellar mass, $M_\ast\sim(1$--$5)\times10^5\,M_\odot$, is considerably smaller than the estimate of its total dynamical mass, $\approx10^7$ to $10^{8}\,M_\odot$\cite{morishita26a}. Assuming that the spatially resolved velocities reflect virial motions, this could indicate a substantial dark matter and/or cold gas component. This finding favors the interpretation of AMORE6 as  an extremely low-mass galaxy, while a star-cluster is dynamically dominated by stellar self-gravity. Finally, the observed line ratio of the Balmer and of the He\,\textsc{ii} $\lambda1640$ and $\lambda4686$ lines are consistent (within the uncertainties) with a negligible amount of dust absorption. 

\begin{figure*}
\centering
	\includegraphics[width=0.99\textwidth]{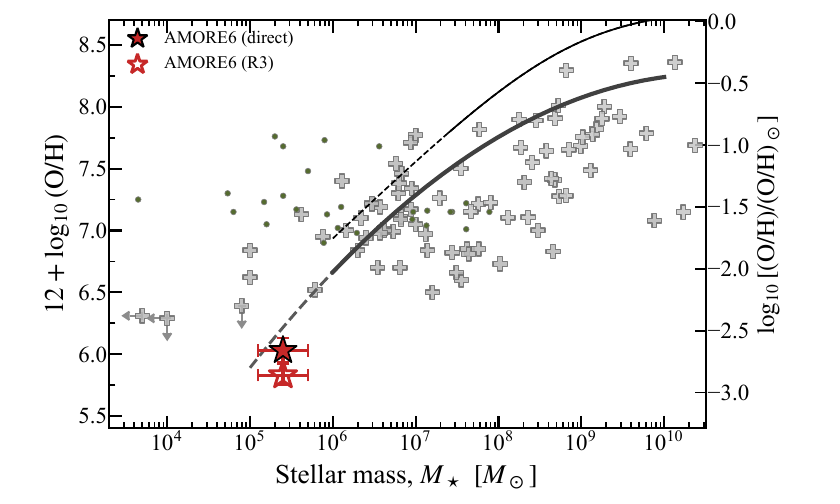}
	\caption{
Stellar mass--metallicity relation (MZR) comparing
the gas-phase oxygen abundance of \id\ with those of
other spectroscopically studied galaxies. The filled red star indicates the oxygen abundance measurement for \id\ via the direct method, with the vertical bar representing the uncertainties in its oxygen abundance and the horizontal error bar indicating the uncertainty in its stellar mass. The open red star indicates the same measurement derived with $R3$ strong line calibration (Methods).
The dark green circles and gray crosses represent metal-poor galaxies at $z\sim0$ \citep{2012A&A...546A.122I,2022ApJ...925..111I} and $z>3$ \citep{vanzella23,nakajima25,chemerynska24,hsiao25,fujimoto25b,maiolino25,willott25,cullen25,mowla24,morishita24b,2026arXiv261105487M}.
The black curves show representative MZR relations, at $z\sim0$ \citep[][thin line]{andrews13} and $\sim6$ \citep[][thick line]{chemerynska24}. The extrapolated mass range is shown with a dashed line respectively.
The right axis indicates the oxygen abundance relative to the solar value.
}
\label{fig:mzr}
\end{figure*}

The estimated oxygen abundance places AMORE6 almost $\sim0.5$\,dex below the extrapolated mass--metallicity relation at $z\sim6$ (Figure~\ref{fig:mzr})\cite{chemerynska24}, indicating unusually limited chemical enrichment even at its low stellar mass. 
Per se, the detection of O and C does not preclude the possibility that AMORE6 is observed during an episode of  star formation in an otherwise pristine environment. PopIII massive stars with stellar masses $M>100M_\odot$ can indeed produce detectable amounts of O and C in very few million years\cite{2026arXiv260921939N}, such that Pop~III massive stars (required to double ionize He) and a low level of chemical   enrichment can co-exist for a few million years. Otherwise, its exceptionally low oxygen and carbon abundances may  indicate that the galaxy has experienced very limited previous star formation, or that its gas has remained largely unaffected by earlier episodes of chemical enrichment. In both cases, the combination of extreme ionization and low metallicity suggests that  a top-heavy IMF is naturally produced in these environments, under the assumption   that the radiation source is of stellar origin.

In conclusion, AMORE6 combines two unusual properties: a remarkably low inferred oxygen abundance and the presence of hard ionizing radiation traced by two \heii\ recombination lines. AMORE6 is consistent with retaining conditions resembling those of primordial star formation almost one billion years after the Big Bang.


\begin{figure}
\centering
	\includegraphics[width=0.49\textwidth]{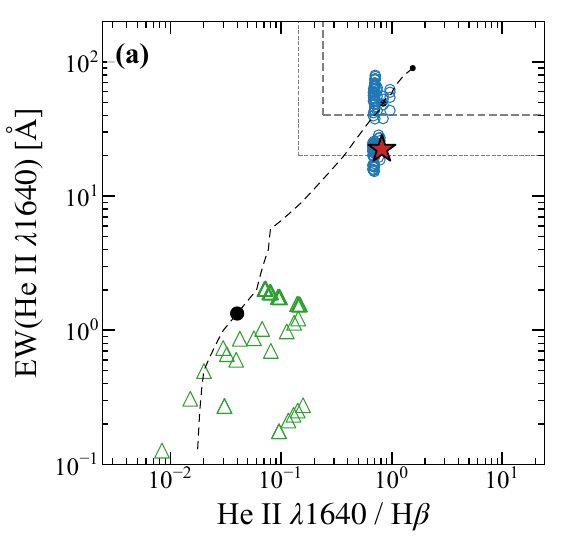}
	\includegraphics[width=0.49\textwidth]{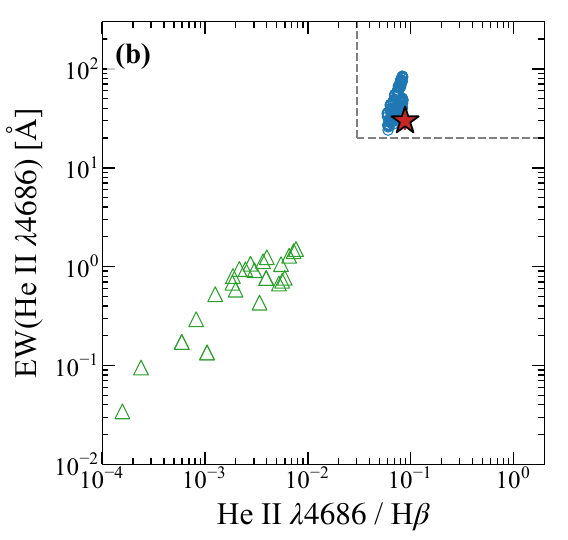}
    \caption{
    Constraints on the \heii\ emission-line strengths of \id\ (red star).
    {\bf (a)} Rest-frame equivalent width of \heiiuv\
    as a function of the \heiiuv\,/\,H$\beta$ flux ratio.  The gray dashed lines indicate the region where Pop-III galaxies---defined as having zero heavy element abundances---are expected to dominate. 
    The dotted lines indicate a relaxed selection criterion that also encompasses AGN\citep{nakajima22}.
    The black dashed line is an evolutionary track for an instantaneous Pop-III burst \citep{schaerer02}. The filled black dots represent characteristic ages, 0, 1, and 2\,Myrs with increasing symbol size.
    Blue open circles and green triangles represent photoionization models calculated for zero metallicity and non-zero metallicity, respectively \citep{morishita26a}.
    {\bf (b)} Same as Panel~(a) but for rest-frame equivalent width of \heii$_{\lambda4686}$
    as a function of the \heii$_{\lambda4686}$/H$\beta$ flux ratio. 
    }
\label{fig:he2}
\end{figure}

\newpage
\clearpage

\section*{Methods}
\label{sec:methods}

Throughout the manuscript, we adopt a standard $\Lambda$CDM cosmology with $\Omega_{\Lambda}=0.7$, $\Omega_{m}=0.3$, and $H_0=70$\,km\,s$^{-1}$\,Mpc$^{-1}$. All magnitudes are given in the AB magnitude system \citep{oke83,fukugita96}. The solar metallicity value $12 + {}$\logoh${}=8.69$\cite{2009ARA&A..47..481A} is adopted throughout this paper.

\subsection*{NIRSpec/MSA observations}\label{sec:nirspec}
We utilize the NIRSpec/MSA observations from GO 9645, ``DAJE-JWST'' (PIs Treu, Fontana, Roberts-Borsani). \id-B was observed as one of the high-priority objects of the program, while \id-A could not be included in the MSA design due to the final pointing and shutter configurations. The observational details and motivation for the program will be presented in a forthcoming survey paper (Treu, T. et al. 2026, in prep.). We provide here a summary of the data reduction for \id-B. 

Briefly, the observations were configured with three medium-resolution ($R\sim1000$) grating configurations (G140M/F100LP, G235M/F170LP, G395M/F290LP), each adopting a 3 shutter configuration, and executed on \dateobs. All exposures used the NRSIRS2 readout. In each visit, all but one nod sequence adopted 19 groups and 7 integrations per exposure (9702\,s on-source), while the remaining nod sequence was shorter, with 19 groups and 2 integrations in visit 003 (2772\,s on-source), 13 groups and 6 integrations in visit 006 (5690\,s on-source), and 16 groups and 6 integrations in visit 008 (7003\,s on-source). The total on-source exposure times for each grating were 50.8\,hrs (G140M/F100LP), 45.2\,hrs (G235M/F170LP) and 38.2\,hrs (G395M/F290LP), for a combined on-source time of 134.2\,hrs. 

We reduced the NIRSpec MSA data using \texttt{msaexp}
(\vermsaexp\cite{brammer22}), beginning with the
uncalibrated exposures downloaded from the Mikulski Archive
for Space Telescopes (MAST).
The reduction utilizes calibration reference files provided
by the JWST Calibration Reference Data System (CRDS),
adopting the pmap context \pmap.
We mitigated the 1/f noise in the individual exposures and
adopted an extended wavelength configuration to recover
spectral coverage beyond the nominal wavelength limits
of each grating.

The individual exposures were processed through standard
calibration procedures, including flat fielding, bar-shadow
correction, wavelength calibration, and flux calibration.
We additionally accounted for wavelength shifts arising
from the offset of the source centroid from the center of
the microshutter along the dispersion direction, using the
source-position-dependent wavelength correction implemented
in \texttt{msaexp} (\texttt{trace\_with\_xpos=True}).
This correction was applied to the individual exposures
prior to their combination to mitigate systematic wavelength
offsets and artificial broadening of emission-line profiles.

The standard path-loss correction was removed, and the
additional path-loss correction implemented in
\texttt{msaexp} was disabled.
Instead, we independently derived a wavelength-dependent
slit-loss correction that explicitly accounts for the
spatially resolved morphology of \id\ and the microshutter
configuration of each exposure (see below).

The 2D spectra were extracted from the individual exposures
and background-subtracted using adjacent shutters.
For each grating, the individual 2D spectra were resampled
onto a common wavelength grid and combined via an
inverse-variance weighted mean.
A 1D spectrum was then extracted from each combined 2D
spectrum using an optimal extraction procedure
\citep{horne86}, making use of the spatial light profile
along the cross-dispersion direction.
To account for the wavelength-dependent broadening of the
spatial profile by the point-spread function (PSF), the
combination of the 2D spectra and the subsequent 1D
extraction were performed independently for each grating.

The spectral coverage of individual gratings is affected
by the gap between the two NIRSpec detectors.
At the redshift of \id, the
\ciii\,$_{\lambda\lambda1907,1909}$ doublet falls within
the detector gap of G140M/F100LP and is therefore not
covered by our observations.
Similarly, \hb\ and the
\oiii$_{\lambda\lambda4960,5008}$ lines fall within the
detector gap of G395M/F290LP.
These rest-frame optical lines are, however, recovered
through the extended wavelength coverage of the
G235M/F170LP observations, allowing their fluxes or upper
limits to be constrained.

Finally, the 1D spectra obtained with the three gratings
were combined onto a composite wavelength grid constructed
from their native wavelength grids, preserving the original
wavelength sampling wherever possible.
In wavelength regions covered by multiple gratings, we
adopted the native grid of the grating with the finer
wavelength sampling, avoiding redundant or partially
overlapping wavelength bins.
The individual spectra were resampled onto this composite
grid using a flux-conserving scheme based on the fractional
overlap of the input and output wavelength bins.
The associated variances were propagated using the squared
fractional overlap coefficients, and the resampled spectra
were combined via an inverse-variance weighted mean.
Only output bins fully covered by valid input pixels were
included in the combination, thereby preventing interpolation
across detector gaps or other masked spectral regions.
The native wavelength grids of the individual gratings
were retained separately for the 2D spectra.

We also identified a set of emission features in the
2D spectra originating from a contaminating source at
$z=6.1858$ (MSAID10466), based on the consistent observed wavelengths
of multiple Balmer and [O\,\textsc{iii}] emission lines (Figure~\ref{fig:spec_all}).
The expected wavelengths of these contaminating lines
are indicated separately in the spectral visualization
to distinguish them from emission associated with \id.

\subsection*{Slit-loss correction and absolute flux calibration}
Given the spatially resolved and asymmetric morphology of \id-B, the standard path-loss correction based on a point-source approximation may not adequately account for the fraction of light entering the NIRSpec microshutters.
We therefore developed a custom wavelength-dependent slit-loss correction based on the intrinsic morphology of the source and the observing configuration of each individual exposure.

We reconstructed the intrinsic surface brightness distribution
of \id\ using the best-fitting morphological model derived
from the NIRCam F115W image, consisting of two S\'ersic
components and one point-source component.
The intrinsic model was convolved with monochromatic,
wavelength-dependent PSFs generated using \texttt{STPSF}.
We adopted the same intrinsic morphology at all wavelengths,
such that the wavelength dependence of the modeled spatial
light distribution arises solely from the PSF.

For each exposure, we reconstructed the positions and
orientations of the open microshutters using the source
coordinates, intra-shutter source positions, shutter
configuration, and aperture position angle recorded in the
FITS headers.
The relative astrometric offset between the nominal source
position and the best-fitting morphological components was
preserved throughout the calculation.
The geometric throughput at each wavelength was then
calculated by integrating the PSF-convolved surface brightness
distribution within the open shutter apertures and dividing
it by the total flux of the model.

The wavelength-dependent slit-loss correction was defined
as the inverse of the geometric throughput.
We adopted the median correction across the individual
exposures and interpolated it onto the wavelength grid of
the combined 1D spectrum.
Both the flux densities and their associated statistical
uncertainties were scaled by the resulting correction
factors.
To assess the sensitivity of the correction to uncertainties
in the source placement within the microshutters, we repeated
the calculation after perturbing the source position by up
to $\pm20$\,mas in the cross-shutter direction.

Our correction accounts for the wavelength-dependent
geometric loss of light at the microshutter apertures,
including the effects of the extended source morphology,
PSF broadening, and intra-shutter source placement.
It does not explicitly model diffraction at the shutter
boundaries or subsequent light propagation through the
NIRSpec optical system.
We therefore use this correction primarily to recover the
relative spectral shape rather than to establish an
independent absolute flux calibration.

Finally, the absolute flux scale of the slit-loss-corrected NIRSpec spectrum was anchored to the H$\beta$ flux independently measured from the NIRCam WFSS\cite{morishita25}.
The resulting multiplicative scale factor of \scalehb\
was applied uniformly across the entire spectrum.
All emission-line fluxes and continuum flux densities
reported in this work are based on this rescaled spectrum.

\subsection*{Emission line fitting}\label{sec:linefit}
We estimated the local continuum in each fitting window using a first-order polynomial, masking spectral pixels within $\pm150$\,\AA\ (observed frame) of the expected wavelengths of emission lines. We then fitted Gaussian line profiles to the emission lines in the continuum-subtracted 1D spectrum.

The peak amplitude and Gaussian width were treated as independent free parameters for each line, while all lines within each fitting window shared a common redshift parameter. The amplitudes were sampled in logarithmic space to ensure non-negative line fluxes. The two components of the C\,{\sc iv} $\lambda\lambda$1548,1551 doublet were constrained to have the same Gaussian width, while both the Gaussian widths and the flux ratio of the [O\,{\sc iii}] $\lambda\lambda$4960,5008 doublet were tied, with the latter fixed to the theoretical ratio of 1:2.98.

We first obtained a best-fitting solution using the Nelder--Mead minimization method and subsequently sampled the posterior distributions of the model parameters using the Markov chain Monte Carlo method implemented in {\sc emcee} \cite{foreman13}, assuming a Gaussian likelihood and uniform priors within the adopted parameter bounds. The walkers were initialized around the best-fitting solution and evolved for 500 steps. The first 250 steps were discarded as burn-in, and every fifth sample of the remaining chain was retained.

For each posterior sample, the integrated flux of each line was estimated by numerically integrating the corresponding Gaussian model within one full width at half maximum (FWHM) on either side of the line centroid. The median of the resulting posterior flux distribution was adopted as the fiducial line flux.

Flux uncertainties were estimated using three approaches: first, from the 16th and 84th percentiles of the posterior flux distribution; second, from the observed flux density uncertainties weighted by the Gaussian line profile; and third, from the observed flux density uncertainties within a fixed velocity interval of 100\,km\,s$^{-1}$ centered on the expected line wavelength. The largest of the three uncertainty estimates was adopted for each emission line.

Integrated line fluxes and continuum fluxes are reported in Table~\ref{tab:lines}. The measured Balmer line ratios, particularly H$\alpha$/H$\beta$, H$\gamma$/H$\beta$, and H$\delta$/H$\beta$, are consistent with the theoretical expectations for Case~B recombination within the measurement uncertainties, consistent with null dust attenuation.
This agreement across a broad wavelength range provides an independent validation of our relative flux calibration, including the wavelength-dependent slit-loss correction described above. Some of the higher-order Balmer lines may be affected by blending with adjacent He\,{\sc i} transitions.

\subsection*{Equivalent width measurements}\label{ssec:ew}
Rest-frame equivalent widths (EWs) provide useful diagnostics of the underlying stellar populations, particularly their ages and ionizing photon production efficiencies. We estimate the local continuum flux density directly from the NIRSpec spectra using the continuum models obtained during the emission-line fitting procedure, excluding wavelength regions around emission lines to avoid contamination. The rest-frame EW is calculated as
\begin{equation}
{\rm EW}_0 = \frac{F_{\rm line}}{f_{\lambda,\rm cont}}\frac{1}{1+z},
\end{equation}
where $F_{\rm line}$ is the observed integrated line flux and $f_{\lambda,\rm cont}$ is the model continuum flux density at the observed wavelength of the line. Both quantities are scaled by the same factor to account for slit losses in the NIRSpec MSA spectroscopy, leaving the resulting EWs unchanged.

We estimate the uncertainty in the local continuum flux density from half the difference between the 16th and 84th percentiles of the distribution of continuum realizations. When the 16th percentile is non-positive, we instead adopt the 84th percentile as the continuum uncertainty. We also impose a floor on the 84th percentile using the local continuum rms. A continuum detection is considered significant when the ratio of the estimated continuum flux density to its uncertainty is at least three. For emission lines with significant detections of both the line and the underlying continuum, we estimate asymmetric EW uncertainties by combining the lower (upper) bound of the line flux with the 84th (16th) percentile of the continuum flux density.

For emission lines with significant line detections but without significant continuum detections, we derive lower limits on the rest-frame EWs by adopting upper limits on the local continuum flux densities. Conversely, when the continuum is significantly detected but the emission line is not, we derive upper limits on the EWs using the corresponding line flux upper limits. No EW constraint is reported when neither the emission line nor the underlying continuum is significantly detected.

The continuum of \id\ is generally faint and remains undetected across much of the NIRSpec wavelength range. However, the continuum is significantly detected near \heiiuv, allowing us to measure its rest-frame EW directly. Similarly, the continuum near \heiiopt\ is detected, providing yielding its EW measurement. Calculated EWs for all lines are reported in Table~\ref{tab:lines}.

Of particular interest are the EWs of the two \heii\ recombination lines, which probe the hardness of the ionizing radiation field. Figure~\ref{fig:he2} compares the measured EWs and their limits, together with the corresponding \heii/\hb\ flux ratios, against predictions for metal-free and metal-enriched stellar populations \cite{schaerer02,nakajima22}. The observed constraints place \id\ in a region associated with exceptionally hard ionizing sources, consistent with predictions for Population III stellar populations (defined as having zero abundance of heavy elements). In particular, the \heiiuv\ measurements fall within the relaxed Pop III selection region, which can also be occupied by AGN. Although these diagnostics do not uniquely identify the nature of the ionizing source, they provide evidence for an unusually hard ionizing radiation field in \id.

\subsection*{Imaging Data}\label{sec:image-jwst}
We used public NIRCam imaging data from UNCOVER Data Release 3 \citep{suess24}. The data are used in NIRSpec slitloss correction and illustration in Figure~\ref{fig:mosaic} and Extended Data Figure~\ref{fig:slit}. We note that the images were not used to obtain photometric fluxes.

\subsection*{Gravitational lens modeling}\label{sec:lens}
The strong lensing model of the galaxy cluster Abell~2744 adopted in this work (hereafter v1.9) was constructed using the recently developed parametric lens modelling software \textsc{Gravity.jl} \cite{lombardi2024} and will be presented by Di Rosa et al.\ (2026, in preparation). Briefly, the model builds on the mass parameterization and observational catalogs\cite{bergamini23b}. The model was updated with the catalogs of cluster members and multiple images identified\cite{merlin2024} and newly available spectroscopy from several \jwst\ programs. 

The lens model includes 178 multiple images from 62 source families spanning the redshift range $1<z<9.8$. Of these, 175 images belonging to 61 families have spectroscopic redshifts, while only three images from one family lack spectroscopic confirmation. This represents an increase of 29 multiple images overall and 54 spectroscopically confirmed additional images relative to the previous model\cite{bergamini23b}, which was adopted in a previous study of the same object\cite{morishita25}.

The model offers an excellent agreement with the observed positions of the double images, with separations of 0.15\,arcsec and 0.09\,arcsec for A and B, respectively. Using this model, we estimated a magnification value 
$\mu_{\rm B}=$\,\magnifberr\ for \idb, consistent with the previous estimate\citep{bergamini23b,morishita25}. The value and the associated uncertainties were derived from 1000 realizations, generated by randomly extracting parameter samples from the model MCMC chains and computing, for each realization, the magnification at the model-predicted positions of the multiple images. The reported values correspond to the 16th, 50th, and 84th percentiles of the resulting magnification distributions.

\subsection*{Inference of Oxygen Abundance}\label{sec:logoh}
Using the measured fluxes of \hb\ and \oiii, we obtained the oxygen abundance, 12+$\log$(O/H), via two different methods. 

Firstly, we estimated the oxygen abundance using the emissivities of collisionally excited oxygen lines relative to hydrogen recombination lines, providing a complementary approach that is less dependent on the assumed spectral shape of the ionizing radiation than the strong-line calibration. The extremely low oxygen abundance inferred in the previous study\cite{morishita25} suggests inefficient metal-line cooling and consequently high electron temperatures ($T_{\rm e}\gtrsim20,000$\,K). Such high temperatures are also consistent with those inferred for similarly metal-poor galaxies in previous studies\cite{nakajima25}.
Although the non-detection of [O\,{\sc iii}]$\lambda4364$ precludes a direct measurement of the electron temperature, its upper limit provides an additional constraint on the allowed temperature range. Motivated by these considerations, we adopt a Gaussian distribution of electron temperatures centred at $T_{\rm e}=25,000$\,K with a standard deviation of $2,500$\,K. The adopted distribution encompasses the physically motivated temperature range of $20,000$--$30,000$\,K within two standard
deviations, while allowing for a broader range of temperatures consistent with the observational upper limit on [O\,{\sc iii}]$\lambda4364$. We propagate the uncertainties in the measured emission-line fluxes and the adopted electron temperature distribution through Monte Carlo realizations.

The adopted distribution encompasses the physically motivated temperature range of $20,000$--$30,000$\,K within two standard
deviations. Its upper $3\sigma$ boundary approximately corresponds to the electron temperature inferred when assuming
an [O\,{\sc iii}]$\lambda4364$ flux of twice the measurement uncertainty, i.e., two-thirds of the reported $3\sigma$ flux upper limit. The adopted temperature distribution is therefore motivated by both the expected thermal conditions of extremely metal-poor gas and the observational constraints on the auroral line.

Given the presence of prominent \heii\ emission, a fraction of oxygen is expected to reside in ionization stages higher than ${\rm O}^{++}$.
We account for these unobserved ionization stages using an ionization correction factor (ICF) inferred from the helium ionic abundances. For our fiducial low-density solution, the inferred helium ionic abundances are ${\rm He}^{+}/{\rm H}^{+}=0.0528^{+0.0073}_{-0.0069}$ and ${\rm He}^{++}/{\rm H}^{+}=0.0100\pm0.0010$, yielding an upward correction of approximately 0.075\,dex to the oxygen abundance. 

The resulting total oxygen abundance is $12+$\,\logoh\,$=$\,\logohdirecterr, corresponding to approximately \oxygensundirect\ of the solar oxygen abundance. Allowing the electron density to vary between $500$ and $10,000\,{\rm cm}^{-3}$ increases the ionization correction to approximately 0.15\,dex, yielding $12+\log({\rm O/H})\simeq6.12$. However, the observed \civ-to-\hb\ ratio favors a lower density (see below).

Secondly, we utilized the strong line calibrator, $R3\equiv$\,\oiiir\,/\,\hb, which offers a simple conversion between the line ratio $R3$ and oxygen abundance. 
We utilized photo-ionization models calculated in the previous study\citep{morishita25}. Briefly, we ran photo-ionization models using {\tt Cloudy} \citep{osterbrock89,chatzikos23} and calculated $R3$ and \logoh\ in various conditions. We calculate the model for different O/H abundance ratios and ionization parameter $\log U$. For hydrogen density, we adopt $n_{\rm H}=1000$\,cm$^{-3}$; however, changing density to 100\,cm$^{-3}$ or 10000\,cm$^{-3}$ would not significantly affect the final result. For the ionizing source, we use {\tt BPASS} \citep{eldridge17,stanway18} binary stellar radiation assuming an instantaneous star formation history at a stellar age of 1\,Myr, upper mass cut of 100\,$M_\odot$, and the Kroupa IMF, of the same metallicity as the gas component. The model grids obtained are comparable to those presented by \citep{nakajima25} in the metallicity range of interest. 

To determine \logoh, we adopted the model of $\log U=-0.5$, the same value as was used in previous studies\cite{nakajima25,morishita25}. This ionization parameter is consistent with the observed \heii/\hei ratio. With this, we obtain $12+$\,\logoh\,$=\oxygenerr$, or \,\oxygensunerr\ of solar, for the observed $R3$. This constraint is compatible with the total O$^{++}$/H$^{+}$ abundance derived from the direct method above, assuming $T_{e} = 20,000-30,000$ K and $n_{e} = 1,000$ cm$^{-3}$. The detection of \heii\ lines suggests strong radiation fields. Adopting a lower value $\log U=-1.5$ instead\citep{2026arXiv260611345I}, which is the lowest acceptable per observational data in the literature, would increase the oxygen abundance but only by $\sim0.25$\,dex. Similarly, the expected contribution of n(O$^{3+}$), which must be present as it has an ionizing potential close to that of \heii\, yields a maximum increase of $\sim 0.1$ dex following the ICF\cite{2017PASP..129d3001P}.

Based on the above analyses, we adopt the oxygen abundance inferred from the direct-method approach under the fiducial low-density conditions, $12+$\,\logoh\,$=$\,\logohdirecterr. The quoted uncertainties reflect the propagation of the emission-line flux uncertainties and the adopted electron temperature distribution through the Monte Carlo realizations, conditional on the assumed electron density.

The oxygen abundance inferred from the $R3$ calibration is lower by approximately 0.23\,dex, a difference comparable to the systematic uncertainty associated with the adopted ionization parameter in the photo-ionization models. Rather than combining the two estimates, we adopt the direct-method abundance as our fiducial value, given its weaker dependence on the assumed spectral shape of the ionizing radiation. The difference between the two estimates provides an indication of the systematic uncertainty associated with the abundance determination, beyond the quoted statistical uncertainties.
Additional systematic uncertainties may arise from the unknown physical conditions and ionization structure of the gas.
Nevertheless, both approaches consistently indicate an extremely low oxygen abundance in \id, well below 1\,\% of the solar value.

\subsection*{Dense gas scenario}
Our deep spectroscopy revealed weak O and C lines. To interpret the absence of strong metal lines, we investigate UV line-to-\hb\ ratios. Among the accessible UV lines, CIV provides the strongest constraint because of its sensitivity to both metallicity and electron density. We show the \civ-to-\hb\ line ratio in Extended Data Figure~\ref{fig:line_ratios} and compare with photoionization model grids\cite{morishita26a}, calculated using {\tt Cloudy} \citep{1998PASP..110..761F,chatzikos23}. Here we adopt a nebular model grid with ionization parameter $\log U=-0.5$ irradiated by an instantaneous burst of metal-free stars, while exploring a range of electron densities and gas metallicities. 

The original analysis of the NIRCam WFSS\cite{morishita25} left open the possibility that the observed weakness of \oiii\ emission could be caused by an extremely high gas density. In such environments, collisional de-excitation can suppress \oiii\ emission even in chemically enriched gas, causing metallicity estimates based solely on the \oiii/\hb\ ratio to be biased toward artificially low values. Distinguishing between these scenarios requires deep rest-frame ultraviolet spectroscopy, as UV metal lines provide complementary constraints on both chemical enrichment and gas properties\cite{mingozzi22}.

The combined constraints from \civ/\hb\ and \oiiir\,/\hb\ (Extended Data Fig.~\ref{fig:line_ratios}) provide a means of distinguishing between the low-metallicity and high-density scenarios. The observed upper limit is consistent with extremely low gas metallicity, whereas models with higher metallicity require exceptionally high electron densities to suppress both emission lines. These results therefore favor an intrinsically metal-poor interstellar medium over collisional de-excitation as the primary explanation for the weak metal emission in \id. We also find that given the \hei\ line detections, models with high density would increase the strength of \hei$_{\lambda7065}$ above their detection limits.

\newpage

\section*{Extended Data}
\label{sec:extended_data}

\captionsetup[figure]{name=Extended Data Figure}
\setcounter{figure}{0}


\begin{figure*}
\centering
	\includegraphics[width=0.98\textwidth]{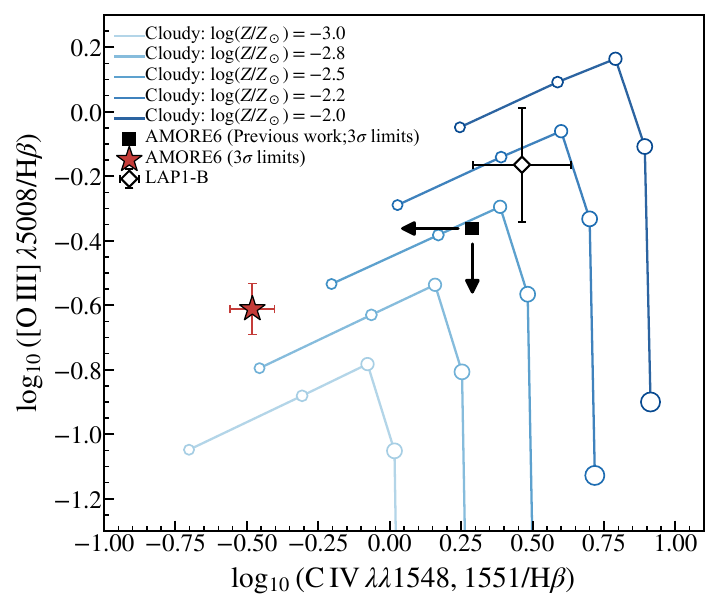}
	\caption{
    The line diagnostic in the \civ/\hb-\oiii/\hb\ space. The red star represents \id, with $1\,\sigma$ uncertainties. For comparison, we show \cl\ photoionization models for different physical parameters --- metallicity $\log Z_{\rm gas}/Z_\odot \in [-3:-2]$ (lines with different colors) and electron density $\log n_e/{\rm cm^{-3}} \in [2:6]$ (circles, size increasing with the density) calculated with $\log U=-0.5$. For comparison, the black diamond is the measurement for LAP1-B\citep[][]{nakajima25}. {The black square and downward arrow indicate the $3\,\sigma$ upper limit on the \oiiir\,/H$\beta$ and \civ\,/H$\beta$ flux ratios of \id\ from previous WFSS and X-shooter observations\citep{morishita26a}.}
    }
\label{fig:line_ratios}
\end{figure*}

\begin{figure*}
\centering
	\includegraphics[width=0.98\textwidth]{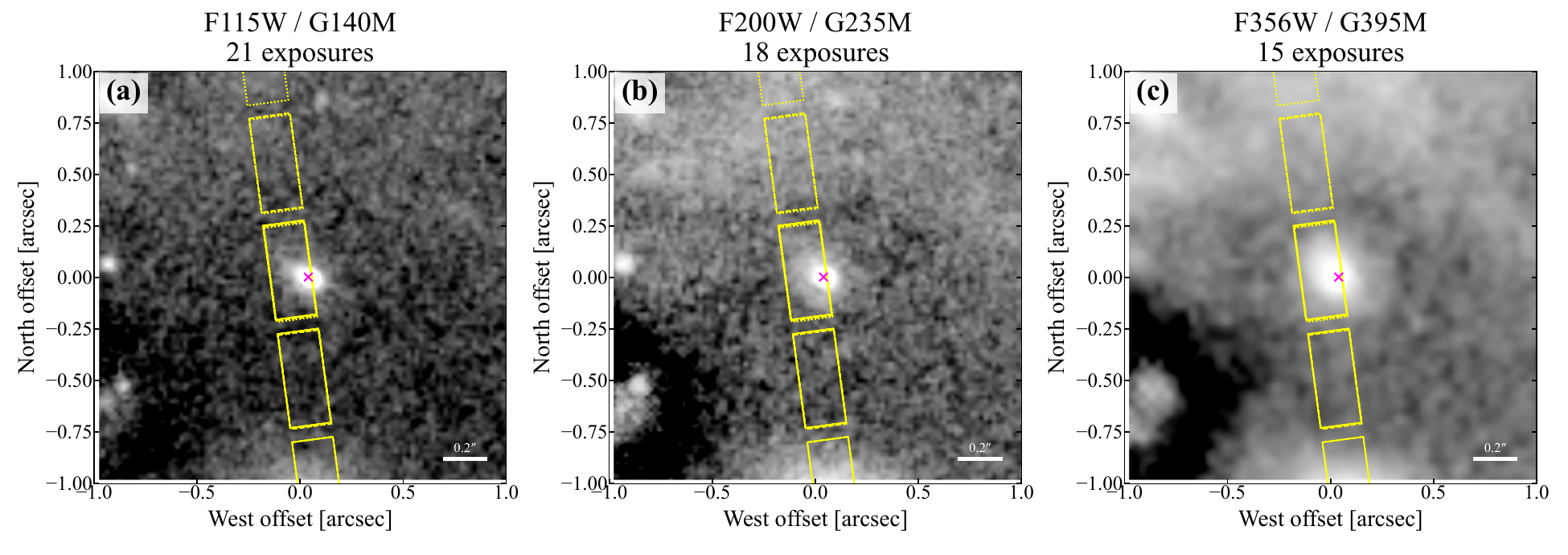}
	\caption{
    Zoomed-in NIRCam images in the F115W ({\bf a}), F200W ({\bf b}), and F356W ({\bf c}) filters, representative of the wavelength ranges covered by the G140M, G235M, and G395M gratings, respectively.
    The yellow rectangles indicate the positions, orientations, and footprints of the NIRSpec MSA microshutters, overlaid for all individual exposures used in the spectroscopic analysis.
    The number of exposures for each grating is indicated above the corresponding panel.
    The magenta crosses mark the source centroid determined from the F115W image and adopted consistently across all three filters.
    These aperture configurations and the wavelength-dependent surface brightness distribution of the source are used to estimate slit losses (see Methods).
    All panels are displayed on the same angular scale, with a $0.\!''2$ scale bar in the lower-right corner of each panel.
    }
\label{fig:slit}
\end{figure*}

\begin{figure*}
\centering
	\includegraphics[width=0.95\textwidth]{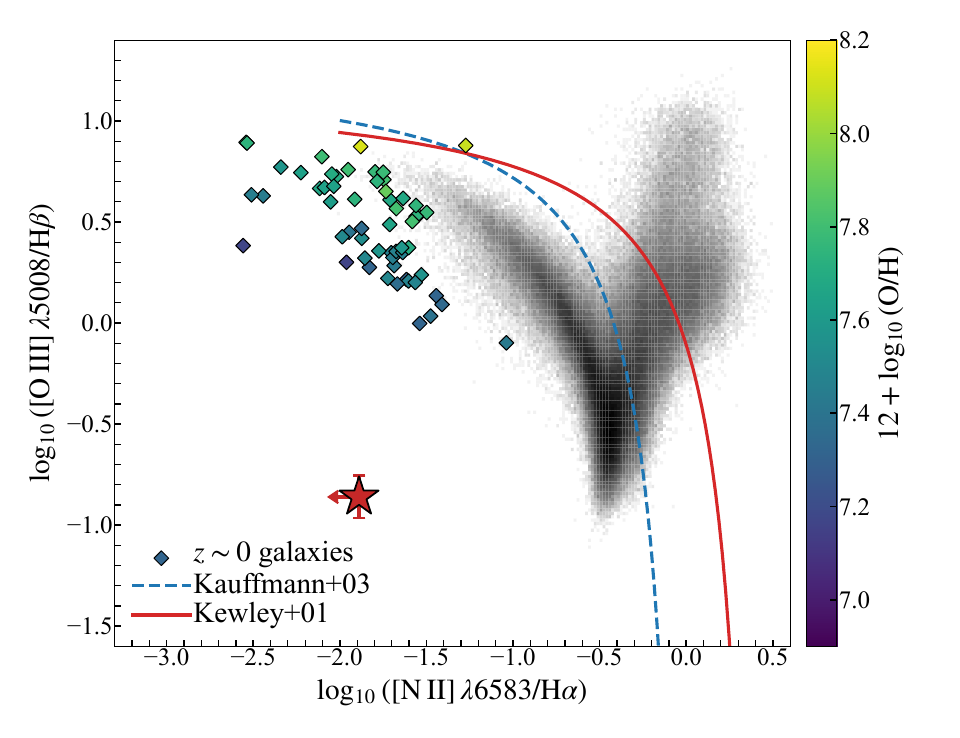}
    \caption{
Baldwin--Phillips--Terlevich (BPT) diagram comparing
the emission-line ratios of \id\ with those of
local galaxies.
The red star indicates \id, with a horizontal arrow representing the $\snlim\,\sigma$ upper limit on \nii$_{\lambda6583}$\,/\ha.
The gray density distribution shows the population
of local galaxies, while the colored diamonds
represent nearby metal-poor galaxies \citep{2012A&A...546A.122I,2022ApJS..262....3N}, color-coded
by their gas-phase oxygen abundance,
$12+\log_{10}({\rm O/H})$.
The blue dashed and red solid curves indicate
the empirical demarcation\cite{kauffmann03}
and the theoretical maximum starburst boundary\cite{2001ApJ...556..121K}, respectively.
The observed low \oiii$_{\lambda5008}$/H$\beta$ and \nii$_{\lambda6583}$/H$\alpha$ ratios place \id\ in an extreme region of the BPT diagram that none of the local galaxies occupy. 
}\label{fig:bpt}
\end{figure*}

\begin{table*}[t]
\centering
\caption{Emission-line fluxes, underlying continuum flux densities, and rest-frame equivalent widths of \id.}
\label{tab:lines}
\small
\setlength{\tabcolsep}{5pt}
\begin{tabular}{lcccc}
\hline
Line & $\lambda_{\mathrm{rest}}$ & Line flux & Continuum flux & Rest-frame EW \\
 & ($\mathrm{\AA}$) & ($10^{-20}\,\mathrm{erg\,s^{-1}\,cm^{-2}}$) & ($10^{-20}\,\mathrm{erg\,s^{-1}\,cm^{-2}\,\mathrm{\AA}}$) & ($\mathrm{\AA}$) \\
\hline
$\mathrm{N\,IV}]$ & 1486 & $<74.6$ & $1.37^{+0.19}_{-0.18}$ & $<8.1$ \\
$\mathrm{C\,IV}$ & 1548 & $44.9 \pm 9.7$ & $1.27^{+0.10}_{-0.10}$ & $5.2 ^{+1.7}_{-1.4}$ \\
$\mathrm{C\,IV}$ & 1551 & $29.8 \pm 9.2$ & $1.27^{+0.10}_{-0.10}$ & $3.5 ^{+1.5}_{-1.3}$ \\
$\mathrm{He\,II}$ & 1640 & $184.1 \pm 7.9$ & $1.22^{+0.06}_{-0.06}$ & $22.4 ^{+2.1}_{-2.0}$ \\
$\mathrm{O\,III}]$ & 1661 & $<12.9$ & $1.22^{+0.06}_{-0.06}$ & $<1.6$ \\
$\mathrm{O\,III}]$ & 1666 & $<13.0$ & $1.22^{+0.06}_{-0.06}$ & $<1.6$ \\
$\mathrm{N\,III}]$ & 1749 & $<9.6$ & $0.93^{+0.06}_{-0.07}$ & $<1.5$ \\
$[\mathrm{O\,III}]$ & 2321 & $<10.0$ & $0.43^{+0.02}_{-0.02}$ & $<3.4$ \\
$[\mathrm{Ne\,IV}]$ & 2423 & $<8.6$ & $0.43^{+0.02}_{-0.02}$ & $<2.9$ \\
$[\mathrm{O\,II}]$ & 3728 & $<4.0$ & $0.15^{+0.01}_{-0.01}$ & $<3.9$ \\
$\mathrm{H}9$ & 3836 & $<12.5$ & $0.15^{+0.01}_{-0.01}$ & $<12.1$ \\
$[\mathrm{Ne\,III}]$ & 3870 & $<11.2$ & $0.15^{+0.01}_{-0.01}$ & $<10.8$ \\
$\mathrm{H}8$ & 3890 & $41.6 \pm 3.2$ & $0.15^{+0.01}_{-0.01}$ & $40.2 ^{+6.9}_{-5.8}$ \\
$[\mathrm{Ne\,III}]$ & 3969 & $<3.7$ & $0.15^{+0.01}_{-0.01}$ & $<3.6$ \\
$\mathrm{H}\eta$ & 3971 & $24.7 \pm 3.0$ & $0.15^{+0.01}_{-0.01}$ & $23.9 ^{+5.2}_{-4.4}$ \\
$\mathrm{He\,I}$ & 4027 & $<2.5$ & $0.11^{+0.02}_{-0.02}$ & $<3.3$ \\
$\mathrm{H}\delta$ & 4103 & $60.2 \pm 5.1$ & $0.11^{+0.02}_{-0.02}$ & $79.0 ^{+28.2}_{-18.6}$ \\
$\mathrm{H}\gamma$ & 4342 & $101.0 \pm 2.8$ & $0.13^{+0.01}_{-0.01}$ & $114.3 ^{+14.7}_{-12.0}$ \\
$[\mathrm{O\,III}]$ & 4364 & $<3.1$ & $0.13^{+0.01}_{-0.01}$ & $<3.5$ \\
$\mathrm{He\,I}$ & 4473 & $<13.8$ & $0.10^{+0.01}_{-0.01}$ & $<20.5$ \\
$\mathrm{C\,III}$ & 4651 & $<2.8$ & $0.10^{+0.01}_{-0.01}$ & $<4.2$ \\
$[\mathrm{Fe\,III}]$ & 4659 & $<1.9$ & $0.10^{+0.01}_{-0.01}$ & $<2.9$ \\
$\mathrm{He\,II}$ & 4687 & $20.0 \pm 2.6$ & $0.10^{+0.01}_{-0.01}$ & $29.8 ^{+6.5}_{-5.5}$ \\
$[\mathrm{Ar\,IV}]$ & 4713 & $<1.9$ & $0.10^{+0.01}_{-0.01}$ & $<2.9$ \\
$\mathrm{He\,I}$ & 4715 & $<2.0$ & $0.10^{+0.01}_{-0.01}$ & $<3.0$ \\
$[\mathrm{Ar\,IV}]$ & 4741 & $<1.9$ & $0.10^{+0.01}_{-0.01}$ & $<2.8$ \\
$\mathrm{H}\beta$ & 4863 & $226.0 \pm 5.0$ & $<0.10$ & $>339.3$ \\
$[\mathrm{O\,III}]$ & 4960 & $18.6 \pm 3.4$ & $<0.40$ & $>7.0$ \\
$[\mathrm{O\,III}]$ & 5008 & $55.3 \pm 10.1$ & $<0.40$ & $>20.7$ \\
$[\mathrm{N\,II}]$ & 5756 & $<1.8$ & $0.05^{+0.02}_{-0.00}$ & $<5.4$ \\
$\mathrm{He\,I}$ & 5877 & $14.5 \pm 1.9$ & $0.05^{+0.02}_{-0.00}$ & $43.7 ^{+10.2}_{-18.4}$ \\
$[\mathrm{S\,III}]$ & 6314 & $<2.2$ & $0.06^{+0.02}_{-0.01}$ & $<5.3$ \\
$\mathrm{Si\,II}$ & 6375 & $<2.1$ & $0.06^{+0.02}_{-0.01}$ & $<5.1$ \\
$[\mathrm{N\,II}]$ & 6550 & $<3.1$ & $<0.07$ & -- \\
$\mathrm{H}\alpha$ & 6565 & $620.2 \pm 5.0$ & $<0.07$ & $>1266.5$ \\
$[\mathrm{N\,II}]$ & 6585 & $<2.9$ & $<0.07$ & -- \\
$\mathrm{He\,I}$ & 6680 & $<3.9$ & $<0.18$ & -- \\
$[\mathrm{S\,II}]$ & 6718 & $<4.1$ & $<0.18$ & -- \\
$[\mathrm{S\,II}]$ & 6733 & $<4.9$ & $<0.18$ & -- \\
$\mathrm{O\,I}$ & 7004 & $<3.4$ & $0.10^{+0.01}_{-0.01}$ & $<5.2$ \\
$\mathrm{He\,I}$ & 7067 & $<3.6$ & $0.10^{+0.01}_{-0.01}$ & $<5.4$ \\
$[\mathrm{Ar\,III}]$ & 7138 & $<3.6$ & $0.10^{+0.01}_{-0.01}$ & $<5.5$ \\
$\mathrm{He\,I}$ & 7283 & $<4.1$ & $<0.47$ & -- \\
$[\mathrm{O\,II}]$ & 7322 & $<4.0$ & $<0.48$ & -- \\
$[\mathrm{O\,II}]$ & 7333 & $<16.1$ & $<0.47$ & -- \\
$[\mathrm{Ni\,II}]$ & 7380 & $<2.9$ & $<0.11$ & -- \\
$[\mathrm{Ni\,II}]$ & 7413 & $<2.7$ & $<0.11$ & -- \\
$[\mathrm{Ar\,III}]$ & 7753 & $<4.3$ & $<0.14$ & -- \\
$\mathrm{O\,I}$ & 7776 & $<4.8$ & $<0.14$ & -- \\
$\mathrm{He\,I}$ & 7818 & $<4.7$ & $<0.14$ & -- \\
\hline
\end{tabular}
\par\vspace{3pt}
\begin{minipage}{0.98\textwidth}
\footnotesize\textit{Note.} 
\textit{Note.} 
All emission-line fluxes and continuum flux densities are scaled by a factor of \scalehb\ to match the H$\beta$ flux independently measured from the NIRCam wide-field slitless spectroscopy (WFSS), accounting for slit losses in the NIRSpec MSA spectroscopy (Methods).
The reported values are not corrected for gravitational magnification.
Quoted uncertainties correspond to $1\sigma$, while upper limits for non-detections are given at $3\sigma$.
Rest-frame equivalent widths are calculated using the locally estimated continuum flux densities. Lower limits on equivalent widths are derived from the $3\sigma$ upper limits on the continuum flux densities, while upper limits are provided for undetected emission lines with detected continua. Dashes indicate cases where neither the emission line nor the underlying continuum is detected, precluding meaningful constraints on the equivalent width.
H8 includes the contribution from He\,{\sc i}\,$\lambda3890$.
H$\beta$ and the [O\,{\sc iii}]\,$\lambda\lambda4960,5008$ doublet are measured in the extended wavelength range of G235M. 
\end{minipage}
\end{table*}


\newpage


\newpage

\section*{Author contributions}
\label{sec:author_contributions}
T.M. led the main analysis, made the figures, contributed to the interpretation of the results, and wrote large parts of the manuscript. 
M.S. carried out an independent data reduction and analysis, and contributed to the interpretation of the results and the writing of the manuscript.
T.T., A.F., and G.R.B designed the observing program, wrote the observing proposal, and selected the targets. G.R.B led the MSA design, with input from T.T. and A.F. Each contributed to the interpretation of the results, and wrote substantial parts of the manuscript.
R.S.R. led one of the data reduction efforts and contributed to the interpretation of the results.
G.D.R., M.L., P.R., L.B., P.B. developed the lens model and computed magnifications.
B.P.-D. helped with the estimation of metallicity.
All authors contributed to the discussion of results and to the final editing of the manuscript.

\section*{Acknowledgements}
 
This work is based on observations made with the NASA/ESA/CSA James Webb Space Telescope. The data were obtained from the Mikulski Archive for Space Telescopes at the Space Telescope Science Institute, which is operated by the Association of Universities for Research in Astronomy, Inc., under NASA contract NAS5-03127 for JWST. These observations are associated with program \#9645. Support for US investigators in program \#9645 was provided by NASA through a grant from the Space Telescope Science Institute, which is operated by the Association of Universities for Research in Astronomy, Inc., under NASA contract NAS5-03127.
TM acknowledges support by JSPS KAKENHI Grant Number 25K24918.
LB and acknowledges the support by the Italian Ministry of University and Research (grant FIS2023-01611, CUP C53C25000300001).
PB acknowledges financial support from the Italian Space Agency (ASI) through contract ``Euclid - Phase E'', and from INAF through the AF2025 grants "PARSE" and "OpenLens-Vet", and the RF2024 grant "Probing Dark Matter and Galaxy Formation in Galaxy Clusters through Strong Gravitational Lensing".
MB acknowledges support from the ERC Grant FIRSTLIGHT \# 101053208, Slovenian national research agency ARIS through grants N1-0238 and P1-0188, and ESA PRODEX Experiment Arrangements No. 4000146646 and 4000149972. 
MC acknowledges support from the INAF GO Grant 2024 ``Revealing the nature of bright galaxies at cosmic dawn with deep JWST spectroscopy''. This work was carried out within the COSMOWEBB project ``A new era for cosmology: exploiting the JWST revolution'', funded by the Italian National Institute for Astrophysics (INAF) through the PRORIS programme (PROgramma di RIcerca Spaziale di base), CUP C83C26000300005. K.G. acknowledges support from Australian Research Council Laureate Fellowship FL180100060. ML acknowledges support from the INAF Curiosity Science Grant AF 2025 "Faint but fierce: stellar outflows driving ionizing photon escape in the early Universe".
EV and MM acknowledge support by 
INAF - Astrofisica di Frontiera 2025, Excellence Track grant ``PARSE'' (PI Vanzella); PRORIS - COSMOWEB ``A new era for cosmology:  exploiting the JWST revolution''. LN acknowledges support from the ERC synergy grant 101166930 - RECAP.
BPD and PS acknowledge financial support from INAF RF2024 Large Grant ``UNDUST: UNveiling the Dawn of the Universe with JWST"
LP acknowledges support from the ERC synergy grant 101166930 - RECAP.
RSR acknowledges support from the Science and Technology Facilities Council (STFC) through a PhD studentship under grant number UKRI1789.
RS acknowledges support from the MUR projects FIS-2024-01621 DAWN and the INFN TEONGRAV initiative.
MT acknowledges support from the ARC Discovery Project DP260100569. 
MM acknowledges financial support from INAF through grants AF2025 ``PARSE" and ``OpenLens-Vet" and RF2024 ``Probing Dark Matter and Galaxy Formation in Galaxy Clusters through Strong Gravitational Lensing".
LG acknowledges support from the MUR projects FIS-2024-01621 DAWN and the INFN TEONGRAV initiative.
MS acknowledges support from the MUR projects FIS-2024-01621 DAWN and the INFN TEONGRAV initiative.
SC acknowledges support from the MUR projects FIS-2024-01621 DAWN.


\section*{Competing interests}
The authors declare no competing interests.

\section*{Data availability}
The JWST NIRSpec observations analysed in this study were obtained  as part of the Deep Amplified JWST Experiment (DAJE; programme ID 9645). The raw observational data are publicly available through  the Mikulski Archive for Space Telescopes (MAST; \url{https://mast.stsci.edu/}). The reduced spectra and derived data products supporting the findings of this study will be made publicly available upon reasonable request or publication.

\section*{Code availability}
The JWST NIRSpec data were reduced using the publicly available  software \texttt{msaexp} (\url{https://github.com/gbrammer/msaexp}). Photoionization models were calculated using the publicly available code \cl\ (ver. 25.00) \citep{1998PASP..110..761F,chatzikos23} (\url{https://www.nublado.org/}).
The custom Python scripts developed for the data reduction,  spectral analysis and physical characterization of the source  are available from the corresponding author upon reasonable request.

\bibliography{manuscript}

\end{document}